# Phase synchronization system receiver module for the Mid-Frequency Square Kilometre Array

Skevos F. E. Karpathakis[a,*], Charles T. Gravestock[a], David R. Gozzard[a,b], Thea R. Pulbrook[a], Sascha. W. Schediwy[a,b]

[a]International Centre for Radio Astronomy Research, The University of Western Australia, 35 Stirling Highway, Perth, Australia, 6009
[b]Australian Research Council Centre of Excellence for Engineered Quantum Systems, The University of Western Australia, 35 Stirling Highway, Perth, Australia, 6009

**Abstract**. Next generation radio telescopes, such as the Square Kilometre Array (SKA) and Next Generation Very Large Array (ngVLA), require precise microwave frequency reference signals to be transmitted over fiber links to each dish to coherently sample astronomical signals. Such telescopes employ phase stabilization systems to suppress the phase noise imparted on the reference signals by environmental perturbations on the links; however, the stabilization systems are bandwidth limited by the round-trip time of light travelling on the fiber links. A phase-locked Receiver Module (RM) is employed on each dish to suppress residual phase noise outside of the round-trip bandwidth. The SKA RM must deliver a 3.96 GHz output signal with 4 MHz of tuning range and less than 100 fs of timing jitter. We present an RM architecture to meet both requirements. Analytical modelling of the RM predicts 30 fs of output jitter when the reference signal is integrated between 1 Hz and 2.8 GHz. The proposed RM was conceived with best practice electromagnetic compatibility in mind, and to meet size, weight and power requirements for the SKA dish indexer. As the ngVLA reference design also incorporates a round-trip phase stabilization system, this RM may be applicable to future ngVLA design.

*Skevos Karpathakis, E-mail: skevos.karpathakis@uwa.edu.au

## 1 Introduction

Radio telescopes of the 21[st] century, such as the Square Kilometre Array (SKA) [1, 2] and Next Generation Very Large Array (ngVLA) [3, 4], require highly precise frequency reference signals at each of the hundreds of antenna sites to enable interferometry and beamforming. The antenna sites in connected element telescope arrays, such as the SKA and ngVLA, may be up to hundreds of kilometers apart; with the optical fiber networks linking these antenna sites being subject to environmental disturbances. These disturbances degrade the precision of the transmitted frequency reference signals and, by extension, the performance of the telescopes [5]. However, optical fiber-based phase stabilization systems can partly compensate for these environmental disturbances and are essential for radio telescope arrays that operate at either high frequencies, or over long optical fiber networks [6-9]. Such telescopes include the Atacama Large Millimetre Array (ALMA), observing at up to 950 GHz [10], and the enhanced Multi-Element Radio Linked Interferometer Network (e-MERLIN) [9], having elements up to 216 km apart. In addition, phase stabilization systems have also been used in very long baseline interferometry (VLBI) observations [11-14].

The Mid Frequency SKA (SKA1-Mid) will operate with a relatively high upper frequency of 15.4 GHz, and with frequency reference signals having to be distributed on aerial fiber spans up to 175 km in length [1], while the Low Frequency SKA (SKA1-Low) will involve exposed ground fibers up to 65 km apart [15]. In the case of SKA1-Mid, previous deployments of reference frequency signals on long-haul optical fiber networks have shown aerial fiber to be up to three orders of magnitude more sensitive to environmental perturbations than buried fiber [16, 17], hence the SKA1-Mid radio telescope has been designed to use an active phase stabilization system. The system works by transmitting the reference signal photonically, generated by a Laser Synthesizer (LS), from a central site to each dish site using separate fiber links. At each antenna site, part of the signal is reflected back along the fiber link to the central site. The phase of this reflected signal is compared to the phase of the transmitted signal, and a phase-locked loop (PLL) is used to suppress any detected phase fluctuations [18-20]. A Transmitter Module (TM) at the central site provides this PLL functionality for each fiber link. However, the PLL feedback system is limited by the light round-trip time (RTT) – the time taken for the signal to travel to the end of the fiber link and back. Fluctuations occurring on the link at time scales shorter than the RTT are unsuppressed by the TM and appear as residual phase noise at frequencies greater than ~300 Hz for the longest links (175 km; fiber refractive index ~1.47).

To account for this residual phase noise, a stable local oscillator and clean-up PLL are included in the Receiver Module (RM) on each dish. The received reference signal is converted from a photonic signal to an electrical signal by a receiver photodetector, and the local oscillator tracks this reference signal with the clean-up PLL. The output signal of the clean-up PLL is used to reference the analog-to-digital converters (ADC) that are used to digitize the astronomical signals received at each dish. In order for the digitization process not to introduce significant timing jitter to the digitized data stream, the local oscillator in the RM must have low broadband phase noise. To maintain maximum correlation efficiency in the presence of in-band radio frequency interference (RFI) at the South Africa site, the effective number of bits (ENOB) required is up to eight bits for two of the five ADC sampling bands [21]. The RM was allocated an integrated timing jitter budget of 100 fs to meet this ENOB requirement in conjunction with the dish ADCs [22].

In addition, another consideration for the SKA1-Mid receiver system is the correlation of radio frequency interference (RFI) signals that are coherent between different pairs of dishes. To reduce this potential source of interference, the SKA1-Mid will employ an offset frequency scheme [23], where unique 1.8 kHz frequency offsets are applied to each dish's reference signal. This sample clock frequency offset (SCFO) scheme requires up to 4 MHz of tuning range above the nominal 3.96 GHz local oscillator frequency. Alternatively, a proportional amount of tuning range is required at a lower frequency, followed by multiplication up to 3.96 GHz. Unfortunately, high tunability and low phase noise are not complementary oscillator properties. No commercial off-the-shelf (COTS) oscillator can satisfy both the SCFO tuning and jitter requirements simultaneously, and therefore a synthesizer architecture is needed.

Furthermore, if the RM is to be co-located on the dish indexer with the ADCs, it must be compact, have low power consumption, and must meet SKA1-Mid electromagnetic compliance (EMC) requirements. To service approximately 200 dishes, the RM must also be cost effective. Therefore, the use of COTS synthesizers as clean-up PLLs is not feasible, and the SKA1-Mid system requires a bespoke module.

This paper presents the design of a SKA1-Mid RM that meets the requirement for low integrated timing jitter and incorporates the SCFO. The RM is designed with best practice EMC in mind, and to be realizable with low size, weight and power. As the SKA1-Mid phase stabilization system was adopted for the current ngVLA reference design [24], this RM may also be suitable for use in a future ngVLA stabilization system.

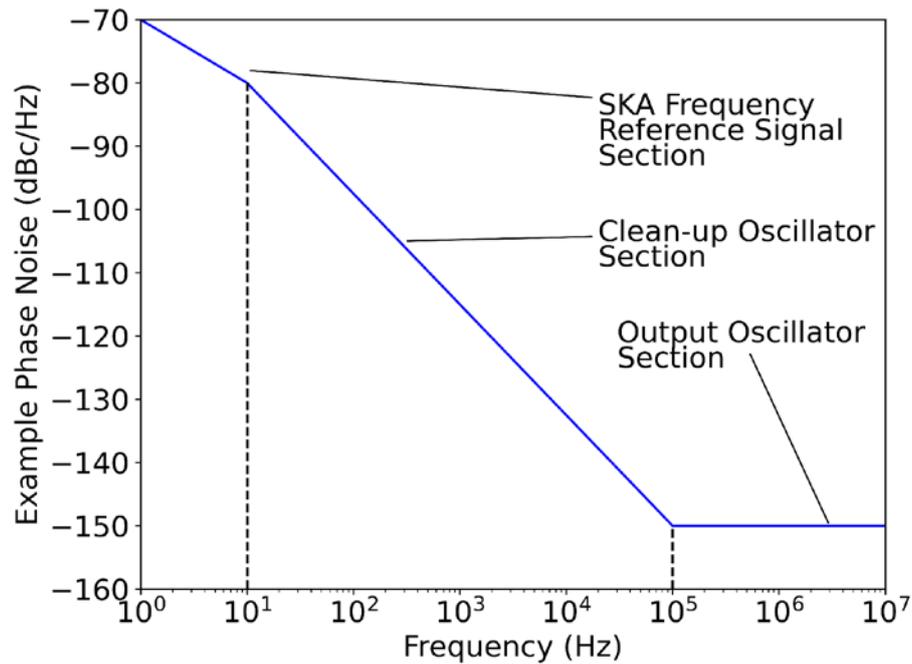

**Fig. 1** Idealized phase noise spectrum with order of magnitude of crossover frequencies indicated. The SKA1-Mid frequency reference signal controls low (<10 Hz) offset frequencies. The high stability local oscillator controls the wideband performance (>100 Hz, <100 kHz). The output oscillator determines the phase noise outside the RM output loop bandwidth (>100 kHz).

## 2 Receiver Module Architecture and Theory

Two RM architectures are considered. A primary architecture meets SKA1-Mid performance requirements by combining a high stability radio frequency (RF) oven-controlled crystal oscillator (OCXO) and a tunable microwave (MW) voltage-controlled oscillator (VCO) in a nested PLL circuit that is phase-locked to the SKA1-Mid frequency reference signal. In this primary architecture, each signal source contributes phase noise to a different band of the RM output phase noise spectrum, as visualized in Fig. 1. A simpler secondary architecture, reflecting the previous RM concept [18], involves a traditional PLL circuit that phase-locks a tunable MW dielectric resonator oscillator (DRO) to the reference signal. A DRO was selected to be the output MW VCO in both architectures because of its low phase noise, relative to other MW VCO families, and COTS availability.

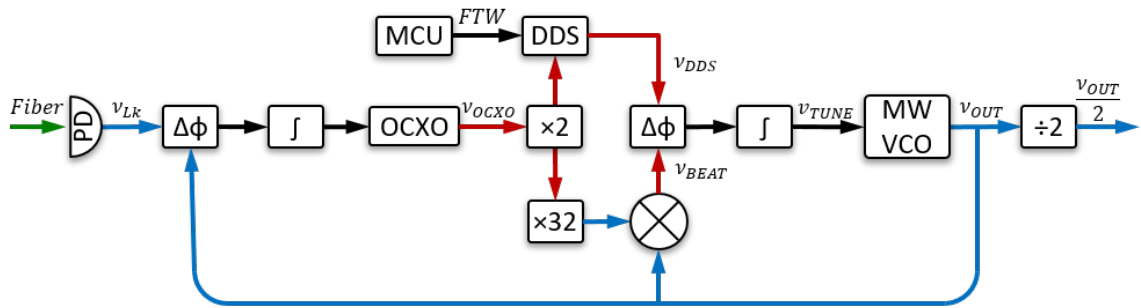

**Fig. 2** Simplified schematic of the proposed receiver module. Edges are labelled with the signal variables defined in this paper. Nodes labelled $\Delta\phi$ are phase detectors or phase-frequency detectors; nodes with integral symbols are integrator circuits. PD, photodetector; Lk, link; OCXO, oven-controlled crystal oscillator; DDS, direct digital synthesizer; MCU, microcontroller unit; FTW, frequency tuning word; MW VCO, microwave voltage-controlled oscillator. The acousto-optic modulator and Faraday mirror [18] are not included in the schematic but do not affect the analysis.

The primary architecture, as presented, benefits from the following design features. The OCXO is temperature-controlled, and is inherently robust against temperature changes at the telescope site. Applying the full range of the SCFO requires an output oscillator with an 8 MHz tuning range, because the frequency reference signal is transmitted at twice the final 3.96 GHz frequency. However, the OCXO is tunable only on the order of parts-per-million of its 125 MHz nominal frequency. Therefore, a direct digital synthesizer (DDS) is used to implement the SCFO, and this complete primary RM architecture is shown in Fig. 2. The DDS has a programmable output that is fine enough to synthesize the 1.8 kHz SCFO steps with the specified 100 Hz resolution [23].

The reference signal received at the photodetector, $v_{Lk}$, has a nominal MW frequency, $v_{MW}$, plus SCFO frequency, $2v_{SCFO}$,

$$\nu_{Lk} = \nu_{MW} + 2\nu_{SCFO}. \qquad (1)$$

This signal is compared with the output VCO by a phase detector (PD). The PD error signal is integrated and tunes the OCXO signal $\nu_{OCXO}$ to track the close-in phase noise of the link, as shown in Fig. 1. The OCXO signal is multiplied and mixed with the RM output signal, $\nu_{OUT}$, producing an intermediate beat frequency, $\nu_{BEAT}$ that is equal to

$$\nu_{BEAT} = \nu_{OUT} - 64\nu_{OCXO}. \qquad (2)$$

The DDS is referenced, and therefore phase-locked, to $2\nu_{OCXO}$. The DDS output signal, $\nu_{DDS}$, is programmed such that

$$\nu_{DDS} = \nu_{MW} - 64\nu_{OCXO} + 2\nu_{SCFO}. \qquad (3)$$

The signals $\nu_{BEAT}$ and $\nu_{DDS}$ are mixed and integrated, producing a tuning voltage, $\nu_{TUNE}$,

$$\nu_{TUNE} = K(\nu_{BEAT} - \nu_{DDS}),$$

$$\nu_{TUNE} = K(\nu_{OUT} - (\nu_{MW} + 2\nu_{SCFO})). \qquad (4)$$

The constant $K$ in equation (4) is the servo gain attributed to the phase detector and integrator circuit. Negative feedback drives $\nu_{TUNE}$, forcing $\nu_{OUT}$ to the same frequency as $\nu_{Lk}$, and suppressing the received signal's residual phase noise. Adjusting the signal $\nu_{DDS}$ to select a different SCFO will force $\nu_{OUT}$ to lock to the new frequency.

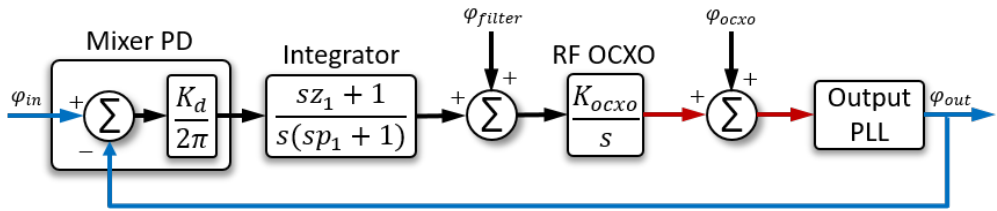

**Fig. 3** Input-connected phase-locked loop (PLL) small signal loop model. PD, phase detector; Kd, detector gain; RF, radio frequency; OCXO, oven-controlled crystal oscillator; Kocxo, OCXO gain.

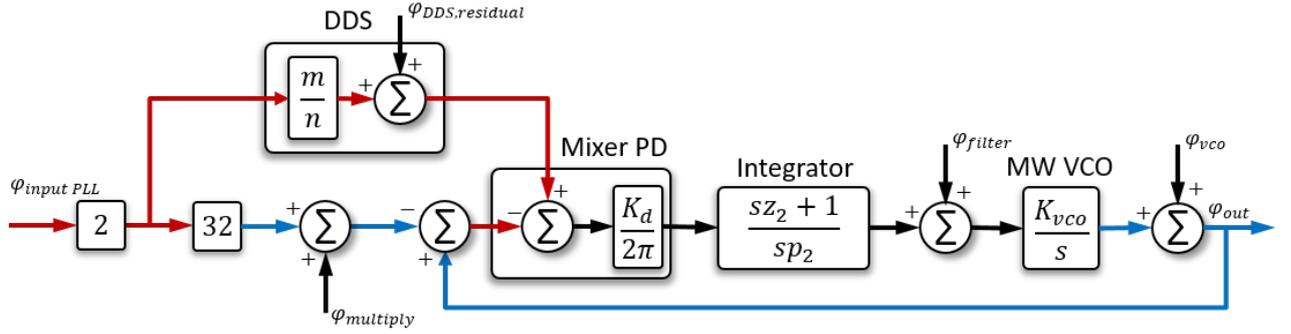

**Fig. 4** Output-connected phase-locked loop (PLL) small signal loop model. DDS, direct digital synthesizer; PD, phase detector; Kd, detector gain; MW, microwave; VCO, voltage-controlled oscillator; Kvco, VCO gain.

## 3 Simulation Method

To evaluate if this design satisfies the SKA1-Mid jitter specification, the phase noise spectrum of the locked RM was simulated by modelling the circuit as a classical phase-domain control system [25]. The RM has a nested PLL structure, with an input loop and output loop. The input loop, seen in Fig. 3, phase-locks the OCXO to the signal $v_{Lk}$ from the photodetector. The output loop, shown in Fig. 4, is connected to the ADCs downstream and phase-locks the output VCO to the OCXO (and adds the signal $v_{SCFO}$).

Optimum jitter performance was achieved by setting the input loop bandwidth at the expected crossover frequency of the OCXO phase noise with the residual phase noise of the fiber link infrastructure. The phase noise of a prototype SKA1-Mid TM, transmitting over 150 km of spooled fiber, was measured in the laboratory. This measurement indicated that an input loop bandwidth on the order of 10 Hz was suitable to meet the output jitter specification. This spooled fiber measurement was used as input data in subsequent simulations to estimate the link induced phase noise. The spooled fiber is an appropriate surrogate for the longest (175 km) span of aerial fiber in SKA1-Mid because the phase noise within the ~300 Hz RTT bandwidth is suppressed by the phase stabilization system regardless of the fiber's packaging, and the 10 Hz PLL bandwidth is within this RTT bandwidth. Above this RTT bandwidth, any phase noise characteristics unique to either spooled or aerial fiber are suppressed by the PLL in the RM.

The complete RM output power spectral density (PSD) is evaluated using Laplace-domain noise transfer functions (NTF) that weight the contribution of input PSDs $S_i(\nu)$ to the output PSD, $S_o(\nu)$,

$$S_o(\nu) = S_i(\nu)|NTF(\nu)|^2. \qquad (5)$$

The $|NTF(\nu)|$ term in Equation (5) is the magnitude of an NTF with $s = j2\pi\nu$. The resultant timing jitter is then an integration of the phase noise spectrum and nominal output frequency, over a single-sided offset frequency domain. In order to perform the above computations, the application GNU Octave was used to evaluate Equation (5), between 1 Hz and 10 GHz. Phase noise spectra for most prototype components were available from the respective manufacturers. However, asymptotic approximations (extending available $f^0, f^{-1}, f^{-2}$ slopes) were necessary for some components because certain noise spectra were not available over the entire frequency domain. The timing jitter is then calculated by integrating phase noise between a lower and upper limit. The 100 fs SKA1-Mid jitter requirement is defined between 10 Hz and 4000 kHz, based on the spectral resolution of planned science cases [22]. Nevertheless, in this analysis, the upper limit was set to correspond with twice the 1.4 GHz bandwidth of the third sampling band of SKA1-Mid, because it is the highest frequency band (1.65-3.05 GHz) requiring eight bits of digitization (the uppermost two frequency bands spanning 3.05-15.4 GHz require four and three bits of digitization, respectively) from the direct-sampling RF ADC. The lower integration limit was set to 1 Hz because the phase stabilization system, without the RM clean-up PLL, has been shown to maintain coherence for integration periods over 1 second [18-20]. Hence the timing jitter results presented are calculated from the phase noise integrated between 1 Hz and 2.8 GHz.

## 4 Simulation results

A phase noise spectrum for the complete RM is shown in Fig. 5. In this simulation the bandwidth of the input and output loops were 18.9 Hz and 1.1 MHz, respectively, and the input loop had 81º of phase margin. The jitter of the SKA1-Mid RM spectrum of Fig. 5 is on the order of 30 fs (integrated between 1 Hz and 2.8 GHz), approximately one-third of the 100 fs SKA1-Mid jitter requirement.

The RM phase noise profile is compared with the secondary single loop phase-locked DRO (PLDRO) architecture in Fig. 6. The PLDRO is locked to the reference signal with a 12 kHz loop bandwidth and has an integrated timing jitter on the order of 128 fs (integrated over 1 Hz to 2.8 GHz, as above). The PLDRO exceeds the required SKA1-Mid jitter by a factor of 1.3 indicating that this concept, initially conceived as an early RM concept [18], is not fit for the current SKA1-Mid requirement. The integrated jitter between 10 Hz and 4000 kHz is approximately the same, because the majority of phase noise is within this band.

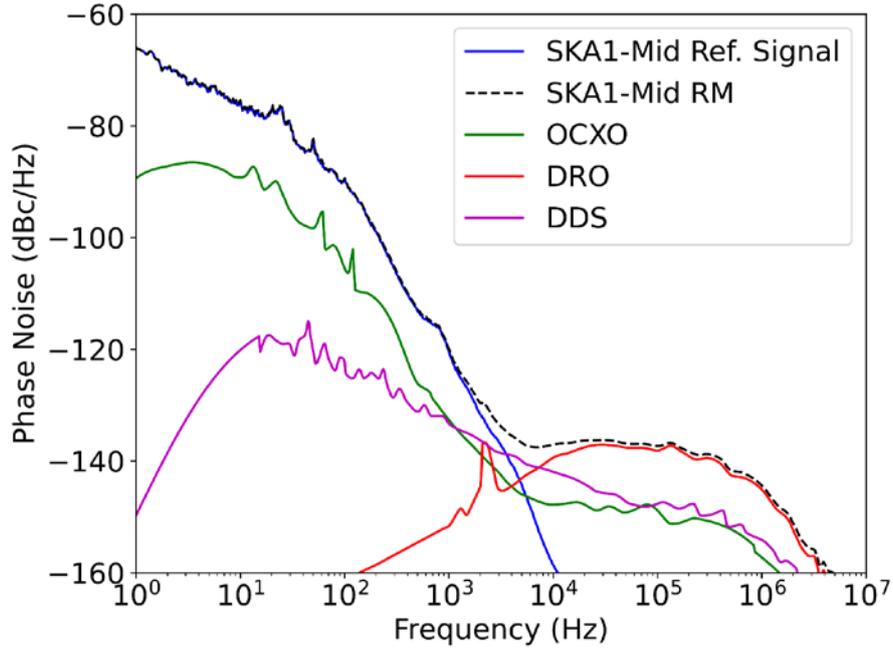

**Fig. 5** Phase noise spectra of key noise contributors at the SKA1-Mid Receiver Module (SKA1-Mid RM) output, and sum, at 8 GHz. All phase noise spectra are scaled to the 8 GHz output frequency within the corresponding input and output loop bandwidths and are suppressed outside these loop bandwidths. Input loop bandwidth is 18.9 Hz with 81° of phase margin, the output loop bandwidth is 1.1 MHz. Ref. Signal, frequency reference signal; OCXO, oven-controlled crystal oscillator; DRO, dielectric resonator oscillator; DDS, direct digital synthesizer.

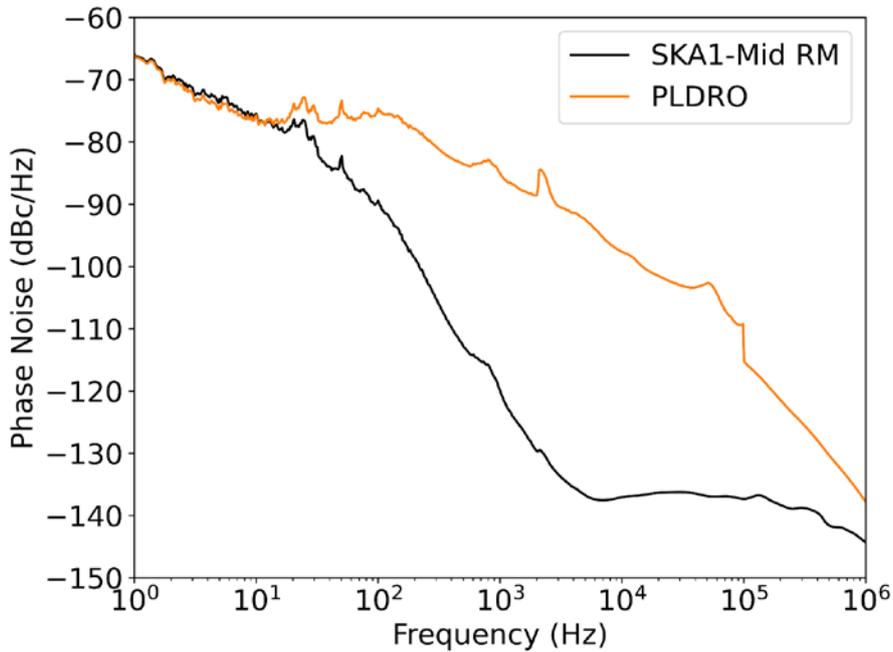

**Fig. 6** Comparison of the SKA1-Mid Receiver Module (SKA1-Mid RM) with a single loop phase-locked dielectric resonator oscillator (PLDRO). Loop bandwidth of the PLDRO is 12 kHz.

# 5 Discussion

## 5.1 Receiver Module Phase Noise

The results shown in Fig. 5 demonstrate the contribution of each signal source to the SKA1-Mid RM phase noise spectrum. Within the 18.9 Hz bandwidth, the reference signal's residual phase noise is unsuppressed and dominates the spectrum. Optimizing the input PLL bandwidth, roll-off and stability is to be solved experimentally by modifying the loop filter components. Between 18.9 Hz and 10 kHz the residual phase noise of the fiber network is suppressed by the input PLL. However, this residual noise is larger than the unsuppressed OCXO phase noise by 5-10 dB. Secondary bumps observed at 20 Hz and 50 Hz are spurious features of the phase noise data measured on the prototype SKA1-Mid phase stabilization hardware. The RM phase noise between 10 kHz and 1 MHz is attributed to the DRO spectra. Above 1.6 MHz, the RM noise floor approaches the white noise floor of the output DRO, approximately −160 dBc/Hz. The OCXO phase noise specification between 100 Hz and 10 kHz may be lifted by approximately 5-10 dB because the residual phase noise from the reference signal exceeds the OCXO over this band. While the final jitter figure is subject to the construction of the RM, predicting jitter within the SKA requirement suggests potential margin to change the OCXO specification.

## 5.2 Single Loop PLDRO

The PLDRO concept was considered as a simpler alternative to the SKA1-Mid RM. As shown in Fig. 6, this PLDRO exhibits more phase noise over the 20 Hz to 1 MHz band compared to the SKA1-Mid RM, and necessarily has an order of magnitude more jitter. The PLDRO does not meet the current SKA1-Mid jitter requirement, but would require a smaller volume and is overall less complex, should the jitter specification for the phase stabilization system be relaxed. Unlike an OCXO, the DRO output frequency changes in response to temperature. This secondary effect adds uncertainty and disadvantages the PLDRO concept compared to the proposed SKA1-Mid RM.

*5.3 Loop Filter Voltage Noise*

The PLL loop filters, based around operational amplifiers, contribute phase noise by frequency-modulating the connected oscillator with the circuit's intrinsic voltage noise (such as resistor Johnson noise). Modelling this behavior is complex because an analytical noise spectrum and transfer function for each possible amplifier circuit is required. Instead, a filter noise budget may be back-calculated from the RM phase noise spectrum of Fig. 5, based on the filter NTF, with an additional margin. Filter voltage noise can be simulated in a SPICE-based application and compared to this noise budget for acceptance. Ultimately this should be experimentally verified, modifying the input PLL filter as required to minimize jitter. Modification could be realized by adding a secondary filtering network (provided the additional poles do not impair stability) or simply replacing the chosen loop filter components.

*5.4 Packaging and Planning for Electromagnetic Compatibility*

The RM physical envelope is the subject of further design, and depends on integration decisions regarding the dish enclosure and sampling electronics. The PLL electronics should be located as close as possible to the ADC clock synthesizers because the electrical connection will be out-of-loop. If necessary for packaging, the passive photonic devices may be mechanically separate from the PLL, being in-loop with the TM servo. From an electronics packaging point-of-view, attention is required to contain RFI from the OCXO multiplication stage, because translating the OCXO frequency requires three multiplier modules and intermediate amplification. The signals in this section of the circuit do not carry a frequency offset, therefore any introduced RFI would be correlated between different dishes. Two intermediate 1 GHz and 4 GHz signals are generated within the RM and are inside the SKA1-Mid observation bandwidth. Other intermediate frequency signals include 125 MHz and 250 MHz, which are all outside of the observation bandwidth. To drive the anti-reflection acousto-optic modulator, an additional 75 MHz signal is generated, but this signal is also outside the observation bandwidth. EMC compliance will be achieved through appropriate board-level zoning of potentially interfering components and two-to-three levels of conductive enclosure around the entire circuit.

*5.5 Offset Frequency and OCXO Digital Supervision*

A microcontroller unit (MCU) on each RM will provide supervisory functions over the SCFO and OCXO, along with interfacing to telescope management hardware, if necessary. For electromagnetic isolation, a digital interface to the RM should be avoided. However, without an interface, the MCU on the RM will not have a priori knowledge of the offset frequency allocated to each link. Offset frequency channel acquisition can be conducted by the MCU by monitoring the OCXO tuning voltage. A mismatch between the RM input and output frequency will drive the input PLL out of the small signal regime, eventually causing the tuning voltage to rail. The MCU may respond by commanding the next adjacent frequency channel from the DDS, until valid PLL lock is detected. Once locked, if a new channel is commanded, the TM may slew the transmitted offset frequency for slow tracking by the RM. Characterizing the channel acquisition and swapping algorithm is the subject of further design and testing. Additionally, in the long term, OCXO aging (typically 2 parts-per-million over 10 years) will change the intermediate comparison frequencies used to achieve lock. The MCU can infer the extent of aging by monitoring the OCXO supply current and PLL tuning voltages, and adjust the nominal DDS frequency to account for this effect. The scheme outlined in this paper therefore improves the deployment lifetime of the RM.

# 6 Conclusion

This paper proposes a RM design solution with 30 fs of timing jitter, falling within the 100 fs SKA jitter budget specification. This analysis assumes the phase noise spectrum of the spooled fiber test setup is an appropriate surrogate for an equivalent length of aerial fiber, in the context of the SKA1-Mid phase stabilization system. This SKA1-Mid RM can be trimmed with the required 1.8 kHz SCFO using a nested PLL, two oscillators and a DDS. The architecture may be realized with entirely COTS electronics to maximize availability. A small signal control model enabled noise simulation of the locked RM. The optimum bandwidths of the input and output PLLs were 18.9 Hz and 1.1 MHz, respectively. A PLDRO concept was simulated for comparison and achieved 128 fs of jitter, demonstrating the necessity of the OCXO, as used in the SKA1-Mid RM.

Electronic and mechanical layout may progress once a final mechanical design and electronic interface of the SKA1-Mid dish digitizer enclosure is available. Work on a prototype RM is progressing and will yield insight on the performance of the proposed architecture. As the final component between the central frequency source and the ADCs, the RM represents the final mile of the SKA1-Mid frequency reference signal transfer system. Both high performance and flexibility in this solution ensure it is fit for the SKA and suitable for the baseline design of the ngVLA frequency reference signal receiver.

*Acknowledgments*

This paper described work undertaken as part of the SKA project. The SKA project is an international effort to build the world's largest radio telescope, led by the SKA Organisation, with the support of 12 member countries. This work was supported by funds from the Australia SKA Office within the Australian Government's Department of Industry, Science, Energy and Resources.

**Skevos F. E. Karpathakis** is an engineer at the International Centre for Radio Astronomy Research. He received his BS in physics and engineering science and MPE in electrical and electronic engineering from The University of Western Australia in 2017 and 2019, respectively. His graduate work involves providing electrical engineering support for radio astronomy research.

**Charles T. Gravestock** is an engineer at the International Centre for Radio Astronomy Research. He received his BS in engineering science and MPE in electrical and electronic engineering from The University of Western Australia in 2016 and 2018, respectively. His current work involves hardware development for radio astronomy research and optical communications research.

**David R. Gozzard** is a Research Associate and Forrest Postdoctoral Fellow at the International Centre for Radio Astronomy Research. He received his BE in mechanical engineering and BS in physics in 2013 from The University of Western Australia. He received his PhD in physics in 2018, also from The University of Western Australia. His current research interests include coherent free-space laser links for metrology and communications, and radio telescope engineering. David is supported by a Fellowship from the Forrest Research Foundation.

**Thea R. Pulbrook** is a Master of Professional Engineering student at The University of Western Australia. She received her Bachelor of Philosophy from the same university in 2018, majoring in electrical engineering and synthetic chemistry, with First Class Honours in chemistry. In 2019 and 2020, she completed a thesis with the International Centre for Radio Astronomy Research as part of the MPE degree.

**Sascha W. Schediwy** is a Senior Research Fellow, and the leader of the Astrophotonics research group at The University of Western Australia. He received his BS in physics with First Class Honours from Curtin University in 2002 and his PhD degree from The University of Western Australia in 2007. He is the author of 55 journal papers with over 3,000 citations. His current research interests include high-precision photonics for radio telescopes, space science and frequency metrology.